\documentclass[aps, prd, preprint, amsfonts,
  amssymb, amsmath, showpacs, letterpaper,nofootinbib,superscriptaddress]{revtex4-1}

\usepackage[T1]{fontenc} % if needed

\usepackage{slashed}
\usepackage{braket}

\newcommand{\be}{\begin{equation}}
\newcommand{\ee}{\end{equation}}
\newcommand{\bea}{\begin{eqnarray}}
\newcommand{\eea}{\end{eqnarray}}

\newcommand{\nn}{\nonumber\\}

\renewcommand{\ol}{\overline}
\usepackage{color}

\begin{document} 

\title{Symmetries and conservation laws in non-Hermitian field theories}

\author{Jean Alexandre}
\email{jean.alexandre@kcl.ac.uk}
\affiliation{Department of Physics, King's College London,\\ 
London WC2R 2LS, United Kingdom}

\author{Peter Millington}
\email{p.millington@nottingham.ac.uk}
\affiliation{School of Physics and Astronomy, University of Nottingham,\\ Nottingham NG7 2RD, United Kingdom}

\author{Dries Seynaeve}
\email{dries.seynaeve@kcl.ac.uk}
\affiliation{Department of Physics, King's College London,\\ 
London WC2R 2LS, United Kingdom}

\begin{abstract}
Anti-Hermitian mass terms are considered, in addition to Hermitian ones, for $\mathcal{PT}$-symmetric complex-scalar and fermionic field theories.
In both cases, the Lagrangian can be written in a manifestly symmetric form in terms of the $\mathcal{PT}$-conjugate variables, allowing for an unambiguous definition 
of the equations of motion. After discussing the resulting constraints on the consistency of the variational procedure, we show that the invariance of a non-Hermitian 
Lagrangian under a continuous symmetry transformation does not imply the existence of a corresponding conserved current. Conserved currents exist, but these are associated 
with transformations under which the Lagrangian is not invariant and which reflect the well-known interpretation of $\mathcal{PT}$-symmetric theories in terms of systems 
with gain and loss. A formal understanding of this unusual feature of non-Hermitian theories requires a careful treatment of Noether's theorem, and we give specific 
examples for illustration.

\end{abstract}	

\maketitle

%%%

\section{Introduction}	

Most extensions of the Standard Model of particle physics keep one ingredient
 untouched: Hermiticity of the Hamiltonian, 
which automatically implies real energy eigenvalues and unitary evolution. Hermiticity is, however, a sufficient but not a necessary condition for such behaviour,
and many examples of consistent non-Hermitian quantum-mechanical systems are known~\cite{Bender,Bender:2008gh}. A real energy spectrum and unitary evolution can 
instead be guaranteed if (i) the Hamiltonian is symmetric under the combined action of the discrete space-time symmetries of parity $\mathcal{P}$ and time reversal 
$\mathcal{T}$, and (ii) the energy eigenstates are simultaneously eigenstates of $\mathcal{PT}$. 
The ability to relax Hermiticity in favour of $\mathcal{PT}$ symmetry makes it possible to construct consistent non-Hermitian generalizations of existing 
quantum field theories, and this could open up new avenues beyond the Standard Model.

$\mathcal{PT}$-symmetric field theories with imaginary interactions were studied in refs.~\cite{scalarPT1,scalarPT2,scalarPT3,scalarPT4}, where analytic continuation in the complex plane was used to define the path integral
over field configurations. In the present work, we consider instead anti-Hermitian mass terms for complex scalars and fermions, in addition to the 
usual Hermitian and Dirac mass terms, respectively. The latter fermionic theory was originally studied in ref.~\cite{BJR}, and further in refs.~\cite{AB,ABM}, but we revisit here 
the corresponding symmetries, providing new insight into the relationship between conserved currents and invariance of the Lagrangian. 

This article emphasises the following features of non-Hermitian field theories. Firstly, the equations of motion can be defined unambiguously only after performing a detailed study of the
discrete symmetries of the non-Hermitian model. Doing so allows the Lagrangian to be written in a manifestly $\mathcal{PT}$-symmetric form in terms of $\mathcal{PT}$-conjugate variables. 
Secondly, the self-consistency of the equations of motion places non-trivial constraints on the action principle of these non-Hermitian theories. Finally, and as a result of the constraints
on the variational procedure, a continuous symmetry of the Lagrangian does not 
imply the existence of a conserved current, requiring a more careful treatment of Noether's theorem and its derivation~\cite{Noether}. There exist conserved currents, but these do not
correspond to continuous transformations under which the Lagrangian is invariant.

The remainder of the article is structured as follows. In section~\ref{sec:scalar}, we study a complex scalar theory with two fields and an anti-Hermitian mass mixing. After outlining 
its discrete 
and continuous symmetry properties, we show that there exists a conserved current that corresponds to a transformation under which the Lagrangian is {\it not} invariant. Moreover, we find 
that this transformation reflects the well-known interpretation of $\mathcal{PT}$-symmetric theories in terms of coupled systems with gain and loss. In order to understand the origin of 
this conserved current, we first discuss the consistency of the variational procedure in section~\ref{sec:variation}, before describing the formal connection between continuous transformations 
and conservation laws for 
non-Hermitian scalar theories in section~\ref{sec:Noether}. As a second example, in section~\ref{sec:fermion}, we extend our discussions to the theory of a single Dirac fermion with a 
parity-violating and anti-Hermitian 
mass term, which allows us to give a physical interpretation to the conserved current by considering the non-unitary map to the corresponding Hermitian theory. Our conclusions are presented in section~\ref{sec:conc}.

%%%

\section{Non-Hermitian scalar model}
\label{sec:scalar}

In order for a non-Hermitian theory to be viable, its constituent degrees of freedom must possess parity ($\mathcal{P}$) and time-reversal 
($\mathcal{T}$) transformations under which the Hamiltonian is $\mathcal{PT}$ symmetric. It follows that the simplest non-Hermitian free scalar theory 
(without tadpoles) must comprise two complex scalar fields that are coupled through a non-Hermitian mass matrix:\footnote{Throughout this work, the 
term ``mass matrix'' is used to refer to the squared mass matrix.} the presence of two fields allows for non-trivial $\mathcal{P}$ transformations, 
and the complex nature of those fields allows for non-trivial $\mathcal{T}$ transformations. Together, these lead to the usual interpretation of viable 
$\mathcal{PT}$-symmetric theories in terms of coupled systems with gain and loss. 

The non-Hermitian scalar theory of interest is described by the following Lagrangian:
\be\label{Lagrangian}
\mathcal{L}\ =\ \begin{pmatrix}
\partial_\nu \phi_{1}^{\star} & \partial_\nu \phi_{2}^{\star}
\end{pmatrix}\begin{pmatrix}
\partial^\nu \phi_{1} \\ \partial^\nu \phi_{2}
\end{pmatrix}\:-\: \begin{pmatrix}
\phi_{1}^{\star} &  \phi_{2}^{\star}
\end{pmatrix} M^2 \begin{pmatrix}
\phi_{1} \\  \phi_{2}
\end{pmatrix}\;,
\ee
where $M^2 \neq [M^{2}]^{\dagger}$. We take the real, non-Hermitian mass matrix
\be\label{massmatrix} 
M^2\ =\ \begin{pmatrix}
m_1^2 & \mu^2 \\ -\,\mu^2 & m_2^2
\end{pmatrix}
\ee  
and will be interested only in cases for which $m_1^2, m_2^2,\mu^2 \geq 0$.

%%%

\subsection{Discrete symmetries}
\label{sec:DiscrSym}

By defining the field doublet
\begin{equation}
\Phi(x)\ \equiv\ \begin{pmatrix} \phi_1(x) \\ \phi_2(x) \end{pmatrix}\;,
\end{equation}
the transformations of the fields under parity and time reversal can be written in the following general forms:
\begin{align}
\mathcal{P}:&\qquad \Phi(t,\mathbf{x})\ \longrightarrow\ \Phi'(t,-\,\mathbf{x})\ =\ P\,\Phi(t,\mathbf{x})\;,\\
\label{CPT2}
\mathcal{T}:&\qquad \Phi(t,\mathbf{x})\ \longrightarrow\ \Phi'(-\,t,\mathbf{x})\ =\ T\,\Phi^\star(t,\mathbf{x})\;,
\end{align}
where $P$ and $T$ are $2\times 2$ matrices. The complex conjugation arising from the $\mathcal{T}$ transformation is a consequence of its anti-linearity.

Note that we only consider here the discrete space-time transformations of the $c$-number fields. In order to find the corresponding operator-level transformations, we must deal 
with the fact that the action of time reversal is found by considering the matrix elements of the field operators, which, for a non-Hermitian theory, rely themselves on an inner-product 
that is defined with respect to the action of time reversal (see, e.g., ref.~\cite{Bender}). This subtlety becomes clear when one tries to apply the standard operator-level $\mathcal{T}$ 
transformations to the non-Hermitian field theories discussed in this article: one will find that they are not even under the combined action of $\mathcal{PT}$~\cite{Peter}.

Choosing $T=+\,\mathbb{I}_2$, the only possible choice of $P$ (up to an overall complex phase) under which the Hamiltonian is $\mathcal{PT}$ symmetric is
\begin{equation}
P\ =\ \begin{pmatrix} 1 & 0 \\ 0 & -\,1\end{pmatrix}\;,
\end{equation}
i.e.~one of the fields transforms as a {\it scalar} ($\phi_1$) and the other as a {\it pseudoscalar} ($\phi_2$). With these transformations, the non-Hermitian mass term in 
eq.~\eqref{Lagrangian} is both $\mathcal{P}$- and $\mathcal{T}$-odd. Notice that we could actually obtain a $\mathcal{PT}$-symmetric field theory by taking any choice of phases 
for which $PT=\mathrm{diag}(1,-1)$. For instance, choosing $P = \mathbb{I}_{2}$ and $T = \mathrm{diag}(1,-1)$, both fields would transform as scalars. 
However, in order to make manifest the interpretation of this $\mathcal{PT}$-symmetric theory in terms of a coupled system with gain and loss, one should take $T=+\,\mathbb{I}_2$.
Indeed, one expects both $\mathcal{P}$ and $\mathcal{T}$ to swap the source for the sink, which is already provided by the complex conjugation involved in the $\mathcal{T}$. This 
swap is then provided in the $\mathcal{P}$ transformation by the pseudo-scalar property of $\phi_2$.

The eigenvalues of the mass matrix are
\be\label{eigenmasses}
M^2_\pm\ =\ \frac{1}{2}(m_1^2+m_2^2)\: \pm\: \frac{1}{2}\sqrt{(m_1^2-m_2^2)^{2}-4\mu^{4}}\;.
\ee
Thus, we obtain a real and non-degenerate mass spectrum provided that the argument of the square root is positive definite, i.e.~when
\be\label{eta}
\eta\ \equiv\ \frac{2\mu^2}{|m_1^2-m_2^2|}\ <\ 1\;.
\ee
Instead, for $\eta>1$, we obtain a complex conjugate pair of eigenvalues, which are not eigenstates of $\mathcal{PT}$, and the $\mathcal{PT}$ symmetry is broken. 
Throughout the remainder of this work, we will 
consider only the region of unbroken $\mathcal{PT}$ symmetry where $\eta<1$.

The unit eigenvectors of the mass matrix are (taking $m_1^2>m_2^2$)
\be\label{eigenvect}
{\bf e}_{+} \ =\ N \begin{pmatrix} \eta \\ \sqrt{1-\eta^2} -1 \end{pmatrix}\;,\qquad 
{\bf e}_{-} \ =\ N \begin{pmatrix} 1-\sqrt{1-\eta^2}  \\ -\,\eta \end{pmatrix}\;.
\ee
Because of the non-Hermitian nature of the mass matrix, these eigenvectors are not orthogonal with respect to Hermitian conjugation:
\begin{equation}
\label{eq:notortho}
{\bf e}_{+}^{\star}\cdot{\bf e}_{-}\ =\ 2N^2\eta\big(1-\sqrt{1-\eta^2}\big)\;,
\end{equation}
except in the Hermitian limit, $\mu\to 0$ ($\eta\to 0$), as one would expect. However, they are orthogonal with respect to $\mathcal{PT}$:
\begin{equation}
{\bf e}_{+}^{\mathcal{PT}}\cdot{\bf e}_{-}\ =\ 0\;.
\end{equation}
Fixing the normalization with respect to $\mathcal{PT}$, i.e.~$\mathbf{e}_{\pm}^{\mathcal{PT}}\cdot\mathbf{e}_{\pm}=1$, we have
\begin{equation}
N\ =\ \Big(2\eta^2-2+2\sqrt{1-\eta^2}\Big)^{-1/2}\;.
\end{equation}
Notice that $N\to \infty$ as $\eta\to 0$, and the Hermitian limit discussed below eq.~\eqref{eq:notortho} must therefore be taken with care. For $\eta<1$, the two eigenvectors are linearly independent, and span a two-dimensional space. At the exceptional point $\eta\to1$, the eigenvalues merge, the eigenvectors 
become degenerate, and two out of the four original degrees of freedom are lost. Such exceptional points are a well-known feature of non-Hermitian matrices, and they occur 
at the boundary between the regions of broken and unbroken $\mathcal{PT}$ symmetry.

Since the eigenvalues of the mass matrix are real and invariant under $\mu^2\to-\,\mu^2$, it is clear that particles and anti-particles are subject to the same dispersion 
relations, and there must therefore exist a definition of charge conjugation $\mathcal{C}$ under which the action of the theory is $\mathcal{CPT}$ invariant. The consistent 
choice for the charge-conjugation properties of the fields is as follows:
\begin{equation}
\mathcal{C}:\qquad \Phi(t,\mathbf{x})\ \longrightarrow\ \Phi^C(t,\mathbf{x})\ =\ C\,\Phi^{\star}(t,\mathbf{x})\;,
\end{equation}
with $C=P$. The Lagrangian [eq.~\eqref{Lagrangian}] is $\mathcal{PT}$- and $\mathcal{CPT}$-even, but it breaks both $\mathcal{CP}$ and $\mathcal{CT}$ symmetries.

%%%

\subsection{Equations of motion}
\label{sec:eoms}

Up to total derivatives, the Lagrangian can now be written in the manifestly $\mathcal{PT}$-symmetric form
\be\label{symmetricform}
{\cal L}\ =\ \Phi^{\ddagger}\begin{pmatrix} -\,\Box-m_1^2 & -\mu^2 \\ -\mu^2 & \Box+m_2^2 \end{pmatrix}\Phi\;,
\ee
where $\Phi^{\ddagger}(x)\equiv[\Phi^{\mathcal{PT}}(x)]^{\mathsf{T}}$ and the superscript $^\mathsf{T}$ denotes the transpose. 
In this form, it is clear that the set of conjugate variables for the 
non-Hermitian theory are $\{\Phi,\Phi^{\ddagger}\}$, rather than $\{\Phi,\Phi^{\dag}\}$, as would be the case for an Hermitian theory. This observation is consistent 
with the fact that Hermitian conjugation is superseded by the combination of $\mathcal{PT}$ transformation and matrix transposition (which we denote collectively by $\ddagger$) in non-Hermitian 
$\mathcal{PT}$-symmetric theories. For completeness, we note that the 
$\mathcal{CP}$ operation coincides with complex conjugation.

From the above discussion, it follows that the equations of motion are determined consistently by either of the equivalent functional variations
\be
\label{eq:vars}
\frac{\delta S}{\delta\Phi^{\ddagger}}\ =\ 0\qquad \text{or}\qquad \left(\frac{\delta S}{\delta\Phi}\right)^{\ddagger} \ =\ 0~\;,
\ee
giving
\begin{subequations}
\label{equamot}
\bea
\Box\,\phi_1\:+\:m_1^2\phi_1\:+\:\mu^2\phi_2& = &0\;,\\
\Box\,\phi_2\:+\:m_2^2\phi_2\:-\:\mu^2\phi_1&=&0\;.
\eea
\end{subequations}
We recall that the equations of motion for an Hermitian theory are obtained by the functional variations
\be
\frac{\delta S}{\delta\Phi^{\dag}}\ =\ 0\qquad \text{or}\qquad \left(\frac{\delta S}{\delta\Phi}\right)^{\!\dag} \ =\ 0~\;,
\ee
and, by comparing with eq.~\eqref{eq:vars}, Hermitian conjugation is again superseded by $\mathcal{PT}$ transformation and matrix transposition for non-Hermitian theories.
Equivalent equations of motion are also obtained for the non-Hermitian theory from the variations
\be
\frac{\delta S}{\delta\Phi^\star}\ =\ 0\qquad \text{or}\qquad \frac{\delta S^\star}{\delta\Phi}\ = \ 0\;,
\ee
Note, however, that we could choose the following equation of motion instead:
\be
\frac{\delta S}{\delta\Phi}\ =\ 0\qquad \text{or}\qquad \frac{\delta S^\star}{\delta\Phi^\star}\ =\ 0\;,
\ee
which would correspond to the change $\mu^2\to-\,\mu^2$. Since physical quantities depend only on $\mu^4$, as can be seen from the eigenmasses [cf.~eq.~\eqref{eigenmasses}], this alternative choice is equivalent to a field redefinition, $\phi_1\to -\,\phi_1$ say. The physical content of the resulting equations of motion is therefore equivalent 
to those in eq.~\eqref{equamot}.

%%%

\subsection{Current conservation}
\label{sec:phicurrent}

In the Hermitian limit, $\mu\to0$, we can quickly convince ourselves that the $U(1)$ currents of the two complex fields are individually conserved:
\begin{subequations}
\label{eq:U1seps}
\bea
j_{1}^\nu &=& i  \left(\phi_1^\star \partial^\nu \phi_1 - \phi_1 \partial^\nu \phi_1^\star\right)\;, \\
j_{2}^\nu &=& i  \left(\phi_2^\star \partial^\nu \phi_2 - \phi_2 \partial^\nu\phi_2^\star\right)\;.
\eea
\end{subequations}
On the other hand, for $\mu^2\ne0$, these currents are not individually conserved, and their divergence can be found from the equations of motion [eq.~\eqref{equamot}]:
\be\label{divergences}
\partial_\nu j_{1}^\nu \ =\  \partial_\nu j_2^\nu\ =\ i\mu^2 \left(\phi_2^\star \phi_1 - \phi_1^\star \phi_2\right)\;,
\ee
such that the conserved current is
\be\label{conservedcurrent}
j^\nu\ \equiv\ j_1^\nu\:-\:j_2^\nu\;.
\ee
This current corresponds to the phase transformations
\begin{subequations}
\label{oppositecharges}
\bea
\phi_{1}(x)\ &\longrightarrow&\ \phi_{1}'(x)\ =\ e^{+i\alpha}\phi_{1}(x)\;,\\
\phi_{2}(x)\ &\longrightarrow&\ \phi_{2}'(x)\ =\ e^{-i\alpha}\phi_{2}(x)\;,
\eea
\end{subequations}
under which the Lagrangian is {\it not} invariant.
As we will see, this is a consequence of the constraints on the consistency of the variational procedure, and the relationship between 
continuous symmetries of the Lagrangian and conservation laws has to be revisited in non-Hermitian theories. 

We note that the two fields carry opposite charges, and 
one field therefore acts as a source and the other as a sink. This interpretation in terms of gain and loss is characteristic of 
$\mathcal{PT}$-symmetric theories~\cite{Bender}: for the present theory, the 
parity and time-reversal transformations both act so as to interchange the source and the sink.

%%%

\section{Conservation laws}

In this section, we explain more formally the above unusual feature, i.e. the existence of a conserved current in the absence of a symmetry. 

\subsection{Variational procedure}
\label{sec:variation}

From the discussion of the equations of motion in section~\ref{sec:eoms}, it is clear that we cannot simultaneously have
\begin{equation}
\label{eq:ELs}
\frac{\delta S}{\delta \Phi^{\ddagger}}\ =\ \frac{\partial \mathcal{L}}{\partial \Phi^{\ddagger}}\:-\:\partial_{\nu}\,\frac{\partial \mathcal{L}}{\partial(\partial_{\nu}\Phi^{\ddagger})}\ 
=\ 0\qquad \text{and}\qquad \frac{\delta S}{\delta \Phi}\ =\ \frac{\partial \mathcal{L}}{\partial \Phi}\:-\:\partial_{\nu}\,\frac{\partial \mathcal{L}}{\partial(\partial_{\nu}\Phi)}\ =\ 0\;,
\end{equation}
except at the trivial point in the solution space $\Phi=\Phi^{\ddagger}=0$. Hence, for this non-Hermitian field theory, only one of the standard Euler-Lagrange equations can be 
non-trivially satisfied.

The full implications of this observation can be illustrated by considering the first variation of the action
\begin{equation}
S\ =\ \int {\rm d}^4x\;{\cal L}(\Phi,\Phi^{\ddagger},\partial_\nu\Phi,\partial_\nu\Phi^{\ddagger})\;,
\end{equation}
which takes the usual form
\begin{align}
\delta S\ &=\ \int\!{\rm d}^4x\bigg[\bigg(\frac{\partial \mathcal{L}}{\partial \Phi}\:-\:\partial_{\nu}\,\frac{\partial \mathcal{L}}{\partial(\partial_{\nu}\Phi)}\bigg)\delta\Phi\:+\:\delta\Phi^{\ddagger}\bigg(\frac{\partial \mathcal{L}}{\partial \Phi^{\ddagger}}\:-\:\partial_{\nu}\,\frac{\partial \mathcal{L}}{\partial(\partial_{\nu}\Phi^{\ddagger})}\bigg)\nonumber\\
&\qquad\:+\:\partial_{\nu}\bigg(\frac{\partial \mathcal{L}}{\partial(\partial_{\nu}\Phi)}\,\delta\Phi\:+\:\delta\Phi^{\ddagger}\,\frac{\partial \mathcal{L}}{\partial(\partial_{\nu}\Phi^{\ddagger})}\bigg)\bigg]\;.
\end{align}
For an Hermitian theory, the principle of least action ($\delta S=0$) immediately yields the Euler-Lagrange equations when we choose boundary conditions for which the surface terms vanish. 

This is not true of the non-Hermitian theory. If we are to have $\delta S=0$, and at the same time support non-trivial solutions ($\Phi \neq 0$), then at least one of the surface terms must yield a finite contribution. Alternatively, we must couple the system to an external source such that we have support off-shell. In the next section, we will describe how these constraints on the consistency of the variational procedure impact the relationship between continuous symmetries and conservation laws for non-Hermitian field theories.

%%%

\subsection{Symmetry and conserved current}
\label{sec:Noether}

For Hermitian theories, the connection between continuous symmetries and conservation laws gives rise to Noether's theorem~\cite{Noether}. This connection is, however, 
modified in the case of non-Hermitian theories.

Under the transformation 
\be
\Phi\ \longrightarrow\ \Phi\:+\:\delta\Phi\;,\qquad \Phi^{\ddagger}\ \longrightarrow\ \Phi^{\ddagger}\:+\:\delta\Phi^{\ddagger}\;,
\ee
the variation of the Lagrangian is
\be
\delta{\cal L}\ =\ \frac{\partial{\cal L}}{\partial\Phi}\,\delta\Phi\:+\:\delta\Phi^{\ddagger}\,\frac{\partial{\cal L}}{\partial\Phi^{\ddagger}}
\:+\:\frac{\partial{\cal L}}{\partial(\partial_\nu\Phi)}\,\partial_\nu(\delta\Phi)\:+\:\partial_\nu(\delta\Phi^{\ddagger})\,\frac{\partial{\cal L}}{\partial(\partial_\nu\Phi^{\ddagger})}\;.
\ee
This variation can also be written as
\be
\delta{\cal L}\ =\ \bigg(\frac{\partial \mathcal{L}}{\partial \Phi}\:-\:\partial_{\nu}\,\frac{\partial \mathcal{L}}{\partial(\partial_{\nu}\Phi)}\bigg)\delta\Phi\:+\:\delta\Phi^{\ddagger}\bigg(\frac{\partial \mathcal{L}}{\partial \Phi^{\ddagger}}\:-\:\partial_{\nu}\,\frac{\partial \mathcal{L}}{\partial(\partial_{\nu}\Phi^{\ddagger})}\bigg)\:+\:\partial_\nu (\delta j^\nu)\;,
\ee
where  
\be\label{deltaj}
\delta j^\nu\ =\ \frac{\partial{\cal L}}{\partial(\partial_\nu\Phi)}\,\delta\Phi\:+\:\delta\Phi^{\ddagger}\,\frac{\partial{\cal L}}{\partial(\partial_\nu\Phi^{\ddagger})}\;.
\ee
The latter current is conserved iff
\be
\label{eq:residual}
\delta{\cal L}\ =\ \bigg(\frac{\partial \mathcal{L}}{\partial \Phi}\:-\:\partial_{\nu}\,\frac{\partial \mathcal{L}}{\partial(\partial_{\nu}\Phi)}\bigg)\delta\Phi\:+\:\delta\Phi^{\ddagger}\bigg(\frac{\partial \mathcal{L}}{\partial \Phi^{\ddagger}}\:-\:\partial_{\nu}\,\frac{\partial \mathcal{L}}{\partial(\partial_{\nu}\Phi^{\ddagger})}\bigg)\;.
\ee

For an Hermitian theory, we can make use of the Euler-Lagrange equations [eq.~\eqref{eq:ELs}] to show that the current is conserved so long as 
$\delta \mathcal{L}=0$. We then obtain the usual statement of Noether's theorem: {\it For every continuous symmetry of a Hermitian Lagrangian, 
there exists a corresponding conserved current.}

For a non-Hermitian theory, the situation is quite different: we saw, in section~\ref{sec:variation}, that both Euler-Lagrange equations cannot simultaneously vanish on-shell. 
As a result, there exists a conserved current only if we can find a continuous transformation under which the non-Hermitian part of the Lagrangian yields $\delta \mathcal{L}\neq 0$ and such that eq.~\eqref{eq:residual} is satisfied.

As an example, let us return to the model in eq.~\eqref{Lagrangian}. Suppose that we define the equations of motion by
\be
\frac{\delta S}{\delta \Phi^{\ddagger}}\ \equiv\ \frac{\partial \mathcal{L}}{\partial \Phi^{\ddagger}}\:-\:\partial_{\nu}\,\frac{\partial \mathcal{L}}{\partial(\partial_{\nu}\Phi^{\ddagger})}\ = \ 0\;,
\ee
as per the discussions in section~\ref{sec:eoms}. There exists a conserved current for any transformation that satisfies
\be
\delta{\cal L}\ =\ \bigg(\frac{\partial \mathcal{L}}{\partial \Phi}\:-\:\partial_{\nu}\,\frac{\partial \mathcal{L}}{\partial(\partial_{\nu}\Phi)}\bigg)\delta\Phi\;,
\ee
and we therefore require
\be\label{condition}
\delta{\cal L}\ =\ 2\mu^2(\phi_2^\star\,\delta\phi_1-\phi_1^\star\,\delta\phi_2)\;.
\ee

As an example, we consider a phase transformation. The condition in eq.~(\ref{condition}) is satisfied and the current in eq.~(\ref{deltaj}) is conserved iff 
\be
\Phi'\ =\ \exp\left[+\,i\alpha \begin{pmatrix} 1 & 0 \\ 0 & -1 \end{pmatrix}\right]\Phi\;,\qquad \Phi^{\ddagger\prime}\ 
=\ \Phi^{\ddagger}\exp\left[-\,i\alpha \begin{pmatrix} 1 & 0 \\ 0 & -1 \end{pmatrix}\right]\;.
\ee
The two complex fields must have opposite charges and transform according to eq.~(\ref{oppositecharges}), as we found in section~\ref{sec:phicurrent}.

%%%

\section{Non-Hermitian fermion model}
\label{sec:fermion}

We now turn our attention to a fermionic model with both an Hermitian mass term $m\ol\psi\psi$ and an anti-Hermitian 
mass term $\mu\ol\psi\gamma^5\psi$. This model was originally introduced in ref.~\cite{BJR} and has Lagrangian
\be\label{Lf}
{\cal L}_f\ =\ \ol\psi\left(i\slashed\partial\:-\:m\:-\:\mu\gamma^5\right)\psi\;.
\ee
The dispersion relation is
\be
E^2\ =\ \mathbf{p}^2\:+\:m^2\:-\:\mu^2\;,
\ee
and the model is $\mathcal{PT}$-symmetric as long as $|\mu|<m$. For $|\mu|>m$, we obtain a complex conjugate pair of eigenmasses, 
and the $\mathcal{PT}$ symmetry is broken.

In ref.~\cite{AB}, a gauged version of this model was studied, providing a non-Hermitian extension of 
quantum electrodynamics, and it was shown to have the following conserved current:
\begin{equation}
\label{eq:current}
j^{\nu}\ =\ \bar{\psi}\gamma^{\nu}\bigg(1+\frac{\mu}{m}\,\gamma^5\bigg)\psi\;,
\end{equation}
in which the relative probability density of left- and right-handed components depends on the ratio $\mu/m$. At the exceptional point $|\mu|= m$,
one of these two components disappears from the spectrum, and the non-Hermitian features of the model thus allow us to continuously suppress one chirality.
We note that a related result can be found in ref.~\cite{Chernodub}, where a non-Hermitian lattice fermionic system was shown to exhibit unequal 
numbers of right- and left-handed fermions.
The gauged model of ref.~\cite{AB} was studied further in ref.~\cite{ABM}, and it was shown that gauge invariance is broken by 
the non-Hermitian mass term but recovered at the exceptional point. A more detailed discussion of the symmetries of this model is given in ref.~\cite{Peter}, and
we revisit here these properties with a new insight from the developments of the previous sections.

%%%

\subsection{Discrete symmetries and equations of motion}

The $\mathcal{P}$ and $\mathcal{T}$ transformations must be such that their combined action leaves the anti-Hermitian mass term invariant, and we first clarify 
the properties of the $c$-number Dirac field under these transformations. The relevant transformations are given by 
\begin{align}
\mathcal{P}:&\qquad \psi(t,\mathbf{x})\ \longrightarrow\ \psi'(t,-\,\mathbf{x})\ = \ P\,\psi(t,\mathbf{x})\;,\nonumber\\&\qquad \ol\psi(t,\mathbf{x})\ 
\longrightarrow\ \ol\psi^{\,\prime}(t,-\,\mathbf{x})\ =\ \ol\psi(t,\mathbf{x})P\;,\\
\mathcal{T}:&\qquad \psi(t,\mathbf{x})\ \longrightarrow\ \psi'(-\,t,\mathbf{x})\ =\ T\,\psi^{\star}(t,\mathbf{x})\;,\nonumber\\&\qquad\ol\psi(t,\mathbf{x})\ 
\longrightarrow\ \ol\psi^{\,\prime}(-\,t,\mathbf{x})\ =\ \ol\psi^{\,\star}(t,\mathbf{x})\,T\;,
 \end{align}
under which the anti-Hermitian mass term is both $\mathcal{P}$ and $\mathcal{T}$ odd. Having noted the subtlety of defining the equivalent operator-level $\mathcal{T}$ 
transformation in section~\ref{sec:DiscrSym}, we remark that one would instead find that the anti-Hermitian mass term is $\mathcal{T}$ {\it even} under a naive application 
of the usual definition of the time-reversal operator in Fock space (appropriate for Hermitian theories)~\cite{Peter}.
  
In four dimensions, the $P$ and $T$ matrices are given by $P=\gamma^0$ and $T=i\gamma^1\gamma^3$, and $\gamma^5=i\gamma^0\gamma^1\gamma^2\gamma^3$. We work throughout in 
the Dirac basis of the gamma matrices. One can then check that the mass term is symmetric under ${\cal PT}$. Indeed,
the $\mathcal{T}$ transformation proceeds as follows:
\bea
\ol\psi(m+\mu\gamma^5)\psi&\to&\ol\psi^{\,\star} T(m+\mu\gamma^5)T\psi^\star\nonumber\\
&=&\ol\psi^{\,\star} (m+\mu\gamma^5)\psi^\star\nn
&=&[\ol\psi (m+\mu\gamma^5)\psi]^\star\nn
&=&\ol\psi (m-\mu\gamma^5)\psi\;,
\eea
where we have used the facts that $\gamma^5$ is real and the anti-Hermitian mass term is imaginary.
A parity transformation then leads back to the original mass term:
\bea
\ol\psi (m-\mu\gamma^5)\psi&\to&\psi^\dagger(m-\mu\gamma^5)\gamma^0\psi\nonumber\\
&=&\ol\psi(m+\mu\gamma^5)\psi\;.
\eea
The $\mathcal{C}$ transformation is defined as
 \bea
 \mathcal{C}:&\quad \psi(t,x)\ \longrightarrow\ \psi^C(t,x)\ =\ C\ol\psi^{\mathsf{T}}(t,x)\;,\nonumber\\
&\quad \ol\psi(t,x)\ \longrightarrow\ \ol\psi^C(t,x)\ =\ \psi^{\mathsf{T}}(t,x)C\;.
 \eea
In four dimensions, $C=i\gamma^2\gamma^0$, and we may quickly verify that that the anti-Hermitian mass term is $\mathcal{C}$ even, and the Hamiltonian 
(and Lagrangian) itself is $\mathcal{CPT}$ symmetric. In order to make the above symmetries manifest in the kinetic part of the Lagrangian, we recall that 
it is convenient to introduce the anti-symmetrized derivative
\begin{equation}
\overset{\leftrightarrow}{\slashed{\partial}}\ \equiv\ \frac{1}{2}\big(\overset{\rightarrow}{\slashed{\partial}}-\overset{\leftarrow}{\slashed{\partial}}\big)
\end{equation}
via the replacement
\begin{equation}
\ol{\psi}i\slashed{\partial}\psi\ \longrightarrow\ \ol{\psi}i\overset{\leftrightarrow}{\slashed{\partial}}\psi\;.
\end{equation}
 
We are now in a position to write the Lagrangian in terms of $\psi$ and its ${\cal PT}$ conjugate $\psi^{\cal PT}(x)=i\gamma^0\gamma^1\gamma^3\psi^\star(x)$. 
Specifically, we can recast the Lagrangian as
\be
{\cal L}_f(x)\ =\ -\,i\psi^\ddagger(x)\gamma^1\gamma^3(i\overset{\leftrightarrow}{\slashed\partial}-m-\mu\gamma^5)\psi(x)\;,
\ee
where $\psi^\ddagger\equiv(\psi^{\cal PT})^\mathsf{T}$. Under the combined action of $\mathcal{P}\mathcal{T}$, the fields transform as follows:
\begin{subequations}
\begin{align}
\mathcal{PT}:\quad &\psi(x)\ \longrightarrow\ \psi'(x')\ =\ \psi^{\mathcal{PT}}(x)\;,\\
&\psi^{\ddagger}(x)\ \longrightarrow\ \psi^{\ddagger\prime}(x')\ =\ -\:\psi^{\mathsf{T}}(x)\;,
\end{align}
\end{subequations}
and the transformation of the Lagrangian is
\begin{align}
\mathcal{PT}:\quad \mathcal{L}_f(x)\ &\to\ \mathcal{L}_f'(x')\nonumber\\& = 
\ -\,i\psi^{\ddagger\prime}(x')\gamma^1\gamma^3(i\overset{\leftrightarrow}{\slashed\partial^{\,\prime}}-m-\mu\gamma^5)\psi'(x')\nonumber\\
&=\ i\psi^{\mathsf{T}}(x)\gamma^1\gamma^3(-\,i\overset{\leftrightarrow}{\slashed\partial}-m-\mu\gamma^5)\psi^{\mathcal{PT}}(x)\nonumber\\&
=\ \Big[-\,i\psi^\ddagger(x)\gamma^1\gamma^3(i\overset{\leftrightarrow}{\slashed\partial}-m-\mu\gamma^5)\psi(x)\Big]^{\mathsf{T}}\;,
\end{align}
where $\partial_{\nu}^{\,\prime}=\partial/\partial x^{\prime\nu}\equiv\partial/\partial(-\,x^{\nu})$. 
The transposition in the final line is irrelevant, since the indices of the $c$-number spinors are traced over.

The equations of motion are obtained by either of the following equivalent variations:
\be
\frac{\delta S_f}{\delta\psi^\ddagger}\ =\ 0\qquad \mbox{or}\qquad \left(\frac{\delta S_f}{\delta\psi}\right)^{\ddagger}\ =\ 0\;.
\ee
Indeed, we have
\bea
\left(\frac{\delta S_f}{\delta\psi}\right)^\mathsf{T}&=&i(-\,i\slashed\partial^\mathsf{T}-m-\mu\gamma^5)\gamma^3\gamma^1\psi^{\cal PT}\nonumber\\
&=&-\,i\gamma^{1}\gamma^{3}(-\,i\slashed\partial-m-\mu\gamma^5)\psi^{\cal PT}\;,
\eea
with
\be
\mathcal{PT}:\quad \left(\frac{\delta S_f}{\delta\psi}\right)^{\mathsf{T}}\ \longrightarrow\ i\gamma^{1}\gamma^{3}(i\slashed\partial-m-\mu\gamma^5)\psi
\ =\ -\:\frac{\delta S_f}{\delta\psi^\ddagger}\;,
\ee
where the minus sign on the rhs is consistent with the definition of left functional variation for anti-commuting fields. 
These equations of motion are actually equivalent to
\be
\frac{\delta S_f}{\delta\ol\psi}\ =\ 0\qquad \mbox{or}\qquad \frac{\delta S_f^\star}{\delta\psi}\ =\ 0\;.
\ee
Alternatively, we could choose the set of equations of motion to be defined by the variations
\be
\frac{\delta S_f}{\delta\psi}\ =\ 0 \qquad \mbox{or}\qquad \frac{\delta S_f^\star}{\delta\ol\psi}\ =\ 0\;,
\ee
which would result in the change $\mu\to-\,\mu$. As with the scalar case, this is without physical implication, since observables depend only on $\mu^2$. 

%%%

\subsection{Continuous symmetries}
\label{sec:continuous}

We now consider the continuous symmetries of the fermionic Lagrangian in eq.~\eqref{Lf}. We revert to writing everything in terms of $\psi$ and its usual Dirac conjugate $\ol{\psi}$ to avoid a proliferation of gamma matrices.

Under the transformation 
\be
\psi\ \longrightarrow\ \psi\:+\:\delta\psi\;,\qquad \ol\psi\ \longrightarrow\ \ol\psi\:+\:\delta\ol\psi\;,
\ee
the variation of the Lagrangian is                                      
\be
\delta{\cal L}_f\ =\ \frac{\partial{\cal L}_f}{\partial\psi}\,\delta\psi\:+\:\delta\ol\psi\,\frac{\partial{\cal L}_f}{\partial\ol\psi}
\:+\:\frac{\partial{\cal L}_f}{\partial(\partial_\nu\psi)}\,\partial_\nu(\delta\psi)\:+\:\partial_\nu(\delta\ol\psi)\,\frac{\partial{\cal L}_f}{\partial(\partial_\nu\ol\psi)}\;.
\ee
This can be written in the form
\be\label{genericdeltaLf}
\delta{\cal L}_f\ =\ \bigg(\frac{\partial{\cal L}_f}{\partial\psi}\:-\:\partial_\nu\,\frac{\partial{\cal L}_f}{\partial(\partial_\nu\psi)}\bigg)\,\delta\psi\:
+\:\delta\ol\psi\,\bigg(\frac{\partial{\cal L}_f}{\partial\ol\psi}\:-\:\partial_\nu\,\frac{\partial{\cal L}_f}{\partial(\partial_\nu\ol\psi)}\bigg)
\:+\:\partial_\nu (\delta j_f^\nu)\;,
\ee
where we have defined the current
\be\label{deltajf}
\delta j_f^\nu\ =\ \frac{\partial{\cal L}_f}{\partial(\partial_\nu\psi)}\,\delta\psi\:+\:\delta\ol\psi\,\frac{\partial{\cal L}_f}{\partial(\partial_\nu\ol\psi)}
\ =\ \frac{i}{2}\left(\ol\psi\gamma^\nu\delta\psi\:-\:\delta\ol\psi\gamma^\nu\psi\right)\;.
\ee
Taking the equations of motion to be those obtained from
\begin{equation}
\frac{\delta S_f}{\delta\ol\psi}\ \equiv\ \frac{\partial{\cal L}_f}{\partial\ol\psi}\:-\:\partial_\nu\,\frac{\partial{\cal L}_f}{\partial(\partial_\nu\ol\psi)}\ =\ 0\;,
\end{equation}
the current in eq.~\eqref{deltajf} is conserved iff
\be\label{conditiondeltaLf}
\delta{\cal L}_f\ =\ \bigg(\frac{\partial{\cal L}_f}{\partial\psi}\:-\:\partial_\nu\frac{\partial{\cal L}_f}{\partial(\partial_\nu\psi)}\bigg)\,\delta\psi\ =\ -\,2\mu\ol\psi\gamma^5\delta\psi\;.
\ee
The phase transformations satisfying the latter conditions are
\begin{subequations}\label{phasealpha}
\bea
\label{phasealpha1}
\psi&\longrightarrow&\psi' \ =\ \exp\Big[+i\alpha\Big(1+\frac{\mu}{m}\gamma^5\Big)\Big]\,\psi\;,\\
\label{phasealpha2}
\ol\psi&\longrightarrow&\ol\psi'\ =\ \ol\psi\,\exp\Big[-i\alpha\Big(1-\frac{\mu}{m}\gamma^5\Big)\Big]\;,
\eea
\end{subequations}
for which the current in eq.~(\ref{deltajf}) is
\be
\delta j_f^\nu\ = \ \alpha\,\ol\psi\,\gamma^\nu\left(1+\frac{\mu}{m}\gamma^5\right)\psi\;,
\ee
consistent with~eq.~\eqref{eq:current} and~ref.~\cite{AB}. The transformations in eq.~\eqref{phasealpha} again reflect the presence of sinks and sources, 
which in this case are the left- and right-chiral components, as detailed in ref.~\cite{ABM}.

\subsection{Non-unitary mapping}
 
It is instructive to consider an alternative derivation of the conserved current for the fermionic non-Hermitian model, based on the construction of a non-unitary map between this model and an Hermitian one for which the current is known and \emph{does} correspond to a symmetry of the Lagrangian. We look for a similarity transformation $B$ such that the fermion $\chi\equiv B\,\psi$ is described by the Hermitian Lagrangian
\be
\mathcal{L}_{\chi}\ =\ \ol\chi(i\slashed\partial-M)\chi\;,
\ee
where $M=\sqrt{m^2-\mu^2}$. The non-unitary matrix $B$ can be found from the Schr\"odinger form of the equation of motion for $\psi$
\be
i\partial_0\psi\ =\ \gamma_0(\vec\gamma\cdot\vec p+m+\mu\gamma^5)\psi\;.
\ee
In terms of $\chi$, this equation reads
\be
i\partial_0\chi\ =\ B\gamma_0(\vec\gamma\cdot\vec p+m+\mu\gamma^5)B^{-1}\chi\;,
\ee
and it is to be identified with
\be
i\partial_0\chi\ =\ \gamma_0(\vec\gamma\cdot\vec p+M)\chi\;.
\ee
It follows that $B$ must satisfy, for any momentum $\vec p$,
\be
B\gamma^0(\vec\gamma\cdot\vec p+m+\mu\gamma^5)B^{-1}\ =\ \gamma^0(\vec\gamma\cdot\vec p+M)\;.
\ee
Once $B$ is determined, we know that the conserved current is $j^\nu_f=\ol\chi\gamma^\nu\chi$, which, when expressed in terms of the original field $\psi$ is 
\be
\label{newcurrent}
j^\nu_f\ =\ \ol\psi\gamma^0B^\dagger\gamma^0\gamma^\nu B\psi\;.
\ee

Given the structure of the equations, we look for $B$ in the form
\be
B\ =\ a+b\gamma^5\;,\qquad B^{-1}\ =\ \frac{a-b\gamma^5}{a^2-b^2}\;,
\ee
since it leaves the kinetic term unchanged, i.e.
\be
B\gamma^0\vec\gamma\cdot\vec p\,B^{-1}=\gamma^0\vec\gamma\cdot\vec p\;.
\ee
The identification of the mass term gives
\be
(a^2+b^2)m\:-\:2ab\mu\:+\:[(a^2+b^2)\mu-2abm]\gamma^5\ =\ (a^2-b^2)M\;, 
\ee
such that
\be
(a^2+b^2)m\:-\:2ab\mu\ =\ (a^2-b^2)M\qquad \mbox{and}\qquad (a^2+b^2)\mu\ =\ 2abm\;.
\ee
We then find
\be
r\ \equiv\ \frac{b^2}{a^2}\ =\ \frac{1-\sqrt{1-\mu^2/m^2}}{1+\sqrt{1-\mu^2/m^2}}\;,
\ee
and the conserved current \eqref{newcurrent} maps to
\bea
j^\nu_f\ &=&\ \overline{\psi}\gamma^\nu(a^2+b^2+2ab\gamma^5)\psi\nonumber\\
&=&\ a^2(1+r)\overline{\psi}\gamma^\nu\bigg(1+\frac{2\sqrt r}{1+r}\gamma^5\bigg)\nonumber\\
&=&\ a^2(1+r)\overline{\psi}\gamma^\nu\bigg(1+\frac{\mu}{m}\gamma^5\bigg)\psi\;,
\eea
Choosing $a^2(1+r)=1$, we immediately recover the conserved current of the non-Hermitian theory found in the previous subsection.

The present derivation of the conserved current has the advantage that it is independent of the variation of the Lagrangian and focuses on the current itself, which is the essential physical feature. We see that the non-unitary map from the Hermitian to non-Hermitian theory effectively introduces an external (field-dependent) source into the continuity equation for the usual fermionic $U(1)$ current, i.e.
\be
\partial_{\nu}(\bar{\chi}\gamma^{\nu}\chi)\ =\ 0\ \longrightarrow\ \partial_{\nu}(\bar{\psi}\gamma^{\nu}\psi)\:+\:\frac{\mu}{m}\,\partial_{\nu}(\bar{\psi}\gamma^{\nu}\gamma^5\psi)\ =\ \partial_{\nu}(\bar{\psi}\gamma^{\nu}\psi)\:-\:J_{\rm ext}\ =\ 0\;.
\ee
Notice that this is in accord with the conclusions of Sec.~\ref{sec:variation} and that $J_{\rm ext}\to 0$ in the Hermitian limit $\mu\to 0$.

%%%

\subsection{Gauged model}

The model in eq.~\eqref{Lf} was coupled to an Abelian gauge field via both vector and axial-vector terms in ref.~\cite{ABM}, giving the Lagrangian
\be\label{gaugedmodel}
{\cal L} \ =\  -\:\frac{1}{4}\,F^{\rho\sigma}F_{\rho\sigma}\:+\:\ol\psi\left[i\slashed\partial-\slashed A(g_V+g_A\gamma^5)-m-\mu\gamma^5 \right]\psi\;,
\ee
where $F_{\rho\sigma}=\partial_{\rho}A_{\sigma}-\partial_{\sigma}A_{\rho}$ is the usual field-strength tensor. In the massless case ($m=\mu=0$), the action is invariant under the combined vector plus axial-vector gauge transformations
\begin{subequations}
\begin{align}
A_\rho\ &\longrightarrow\ A_{\rho}'\ =\ A_\rho-\partial_\rho\theta\;,\\
\psi\ &\longrightarrow\ \psi'\ =\ \exp\left[i\left(g_V\:+\:g_A\gamma^5\right)\theta\,\right]\psi\;,\\
\ol\psi\ &\longrightarrow\ \ol\psi^{\,\prime}\ =\ \ol\psi\,\exp\left[i\left(-\:g_V\:+\:g_A\gamma^5\right)\theta\,\right]\;.
\end{align}
\end{subequations}
Whilst this gauge invariance is lost in the massive case ($m\ne0$ and/or $\mu\ne0$), it was shown in ref.~\cite{ABM} that the full vector plus axial-vector symmetry is restored at the exceptional point $|\mu|=m$. In this section, we revisit this behaviour in light of the results of section~\ref{sec:continuous}.

To this end, we make the following replacement of the phase $\alpha$ that appears in the global $U(1)$ transformation in eq.~\eqref{phasealpha1} (with a consistent replacement in eq.~\eqref{phasealpha2} for the transformation of the Dirac-conjugate field):
\be
\alpha\ \longrightarrow\ \big(g_V+g_A\gamma^5\big)\theta(x)\;.
\ee
The vector plus axial-vector gauge transformations of the fermion fields then take the form
\begin{subequations}
\begin{gather}
\psi\ \longrightarrow\ \psi'\ =\ \exp\Big[i\big(g_V+g_A\gamma^5\big)\Big(1+\frac{\mu}{m}\gamma^5\Big)\theta\Big]\,\psi\;,\\[0.5em]
\ol{\psi}\ \longrightarrow\ \ol{\psi}^{\,\prime}\ =\ \ol{\psi}~\exp\Big[i\big(-g_V+g_A\gamma^5\big)\Big(1-\frac{\mu}{m}\gamma^5\Big)\theta\Big]\;.
\end{gather}
\end{subequations}
Under these transformations, the mass terms yield a contribution
\begin{equation}
\delta\mathcal{L}\ \supset\ -\,2i\,\theta\,\ol\psi\,\gamma^5
\bigg[\;\mu\,g_V\bigg(1+\frac{\mu}{m}\gamma^5\bigg)\:+\:m\,g_A\gamma^5\bigg(\gamma^5+\frac{\mu}{m}\bigg)\bigg]\psi\;.
\end{equation}
which is equal to $-\:2\,\mu\,\bar{\psi}\,\gamma^5\,\delta\psi$, as required by eq.~\eqref{conditiondeltaLf} for the existence of a conserved current, 
only when $\mu = \pm\,m$. Whilst this is compelling, we must first deal carefully with the additional term that arises from the kinetic term:
\begin{equation}
\label{kineticcontrib}
\delta\mathcal{L}\ \supset\ -\:\bar{\psi}\,\slashed{\partial}\theta\,\big(g_V+g_A\gamma^5\big)\bigg(1+\frac{\mu}{m}\gamma^5\bigg)\psi\;.
\end{equation}
It would appear that this cannot be absorbed via the transformation of the gauge field
\begin{equation}
A_{\rho}\ \longrightarrow\ A_{\rho}'\ =\ A_{\rho}\:-\:\partial_{\rho}\theta\;.
\end{equation}
However, in the limit $\mu=\pm\,m$, the contribution in eq.~\eqref{kineticcontrib} becomes
\begin{subequations}
\bea
-\,(g_V+g_A)\psi_R^{\dag}(2\slashed\partial\theta)\psi_R\quad &\mbox{if}&\quad \mu\ =\ +\,m\;,\\
-\,(g_V-g_A)\psi_L^{\dag}(2\slashed\partial\theta)\psi_L\quad &\mbox{if}&\quad \mu\ =\ -\,m\;,
\eea
\end{subequations}
such that the additional contribution from the kinetic term can be removed by the following transformation of the gauge field:
\begin{equation}
A_{\rho}\ \longrightarrow\ A_{\rho}'\ =\ A_{\rho}\:-\:2\partial_{\rho}\theta\;.
\end{equation}
In this way, we find that that the full vector plus axial-vector symmetry 
is indeed restored in the limit $\mu=\pm\,m$, as found in ref.~\cite{ABM}.

%%%

\section{Conclusions}
\label{sec:conc}

In the context of both scalar and fermionic theories with anti-Hermitian mass terms, we have described the implications of defining self-consistent equations of motion 
for the action principle of non-Hermitian field theories. The resulting constraints on the variational procedure lead to a modification of the usual direct relationship between 
continuous symmetries of the Lagrangian and conservation laws. Most strikingly, in order to find
conservation laws, we are forced to consider transformations that do not leave the non-Hermitian part of the Lagrangian invariant. Whilst this is perhaps surprising, we have shown, for the fermionic model, that the conserved current of the non-Hermitian theory is related to the conserved current of the corresponding Hermitian theory by a non-unitary map. The relevant symmetry transformations of the non-Hermitian Lagrangian appear to 
reflect the well-known interpretation of $\mathcal{PT}$-symmetric theories in terms of coupled systems with gain and loss, and the implications of these observations for model 
building in non-Hermitian theories is a promising avenue to be explored.

As a closing remark, we note that the present work has considered only internal continuous symmetries. We can, however, quickly convince ourselves that, despite the invariance 
of the Lagrangians considered under Poincar\'{e} transformations, neither the standard energy-momentum tensor nor the standard four-dimensional angular momentum current of 
these models are conserved. Even so, one may give a physical interpretation to this result by considering the symmetry transformations of the corresponding Hermitian theory: the non-unitary map from the Hermitian to non-Hermitian theory effectively introduces an external (field-dependent) source into the usual continuity equation that would be anticipated from the Hermitian limit of the theory.

%%%

\acknowledgments

The work of PM is supported by STFC Grant No.~ST/L000393/1 and a Leverhulme Trust Research Leadership Award.

%%%

\end{document}